\documentclass{article}
\usepackage{amsmath, amssymb, amsthm, graphicx,color}
 \DeclareGraphicsExtensions{.pdf,.jpeg,.png}
 \usepackage[multidot]{grffile}
 \usepackage{color}
 \usepackage{subcaption}

\def\etal{\text{\it et al.~}}

\title{Joint denoising and HDR for RAW video sequences}
\date{}
\author{A. Buades \and O. Martorell \and M. S\'{a}nchez-Beeckman}

\begin{document}
%
\maketitle
\begin{abstract}
We propose a patch-based method for the simultaneous  denoising and fusion of a sequence of  RAW multi-exposed images. 
A spatio-temporal criterion is used to select similar patches along the sequence, and a weighted principal component analysis permits to both denoise and fuse the multi exposed data.
The overall strategy permits to denoise and fuse the set of images without the need of recovering each denoised image  in the multi-exposure set, leading to a very efficient procedure. 
Several experiments show that the proposed method permits to obtain state-of-the-art fusion results with real RAW data.
\end{abstract}

\graphicspath{{./figures/}}

\section{Introduction}

Conventional cameras are not able to capture in one shot the whole  range of natural light present in a scene. 
As a consequence, dark or bright regions might appear completely saturated.  A High Dynamic Range (HDR) image has a range of luminosity between the brightest area and the darkest area larger than usual. 
High Dynamic Range imaging refers to the set of methods and techniques that permit to increase the dynamic range of images and videos. 
Particularly, we will deal with the combination of several low dynamic range images of the same scene acquired with different exposure times, in order to create a HDR image.

Most methods combine the radiance values instead of the usual color values of standard 8 bit per pixel and channel images. 
In order to do so, the camera response function (CRF) has to be estimated, generally  using the method proposed by Devebec and Malik \cite{debevec2008recovering}. When using the radiance, the result obtained by a HDR method has to be converted into a standard representation, for visualization in common displays. This process is known as tone-mapping \cite{Reinhard_2005}.  Methods combining the usual color values are referred as Multi Exposure Fusion (MEF) methods \cite{mertens} and do not need additional tone-mapping.

\medskip

\begin{figure}[ht]
\centering
\includegraphics[width=7cm]{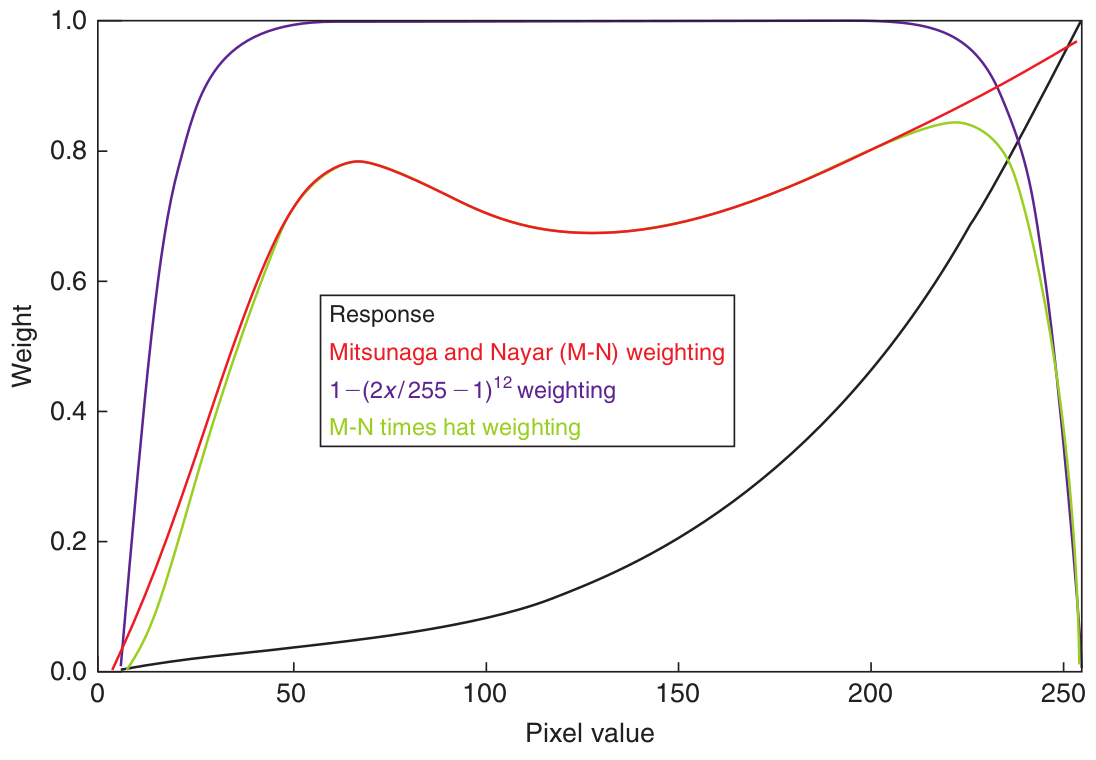}
\caption{Several HDR weighting funcions.}
\label{fig:hdr_weights}
\end{figure}

There exists a wide literature on HDR imaging.  
Many classical methods combine the set of images using a weighted average of irradiance values:
\begin{equation}
E(x,y) = \sum_{i=1}^{N} w(I_i(x,y)) \left( \frac{g^{-1}(I_i(x,y))} { t_i }\right) / \sum_{i=1}^{N} w(I_i(x,y)) 
\label{eq:fusion_hdr}
\end{equation}
where $I_1, \ldots, I_N$ denote the input images, $g$ is the camera transfer function and $w(\cdot)$ is designed to diminish the contribution of under and over exposed pixels. For computing $w(\cdot)$, the most commonly used methods are the ones proposed by Devebec and Malik \cite{debevec2008recovering}, Mitsunaga and Nayar \cite{mitsunaga1999radiometric} or Mann and Picard \cite{mann1994beingundigital}. Figure \ref{fig:hdr_weights} displays some examples of weighting functions. 

In order to apply the weighted average in Equation \eqref{eq:fusion_hdr}, the image sequence needs to be static, that is, the objects in the scene do not move and the camera is fixed. This is not the case for many sequences, and the application of such methods introduces ghosting effects. To avoid these artefacts, methods have to be adapted to non-static image sequences \cite{hu2013hdr, khan2006ghost}. Recently, deep learning based methods have been proposed dealing implicitly with the ghosting effect as a part of the proposed network \cite{kalantari2017deep}.

\medskip

While the application of the inverse transfer function permits to combine values acquired with different exposure, it is not adequate for taking care of noise.  
Most common cameras use a CCD or CMOS sensor device measuring a single color per pixel. 
The selected configuration of the sensor usually follows the Bayer color filter array (CFA) \cite{Bayer}. 
Thus, out of a group of four pixels, two are green (in quincunx), one is red and one is blue \cite{AlleyssonSusstrunkHerault}.  
Demosaicking is the interpolation process by which the two missing color values are estimated. 
The demosaicking is usually performed by combining close values from the same channel or the other two. 
As a result, the noise, being almost white at the sensor, gets color and spatial correlated. 
The rest of the imaging chain, consisting mainly in color and gamma corrections and compression, enhances the noise in dark parts of the image leading to contrasted colored spots of several pixels. 
The size of these spots depends on the applied demosaicking method.  
The inversion of the transfer function is not able to revert all noise correlating stages. 
For this reason, we will perform denoising and HDR fusion in the RAW domain.

\medskip

HDR methods using the weighted average \eqref{eq:fusion_hdr} actually reduce noise, but this reduction is rather minor due to the small number of values being involved. 
Video denoising methods take into account a spatio-temporal neighborhood of each pixel under consideration improving the noise reduction capabilities. 
Inspired by the video denoising method in \cite{vdenoisingTIP15} and classical HDR methods, we will propose a joint denoising and HDR method removing noise and building a HDR image free of ghosting artifacts.

\medskip

The paper is organized as follows:
Section \ref{sec:related} describes the existing literature on MEF methods. 
In Section \ref{sec:method} we describe the complete method.
In Section \ref{sec:experimentation}, we discuss the implementation of the method and compare with state of the art algorithms. 

\section{Related work} \label{sec:related}

\subsection{Image sequence white noise removal}

Local average methods, as the bilateral filter \cite{tomasi1998bilateral},  or patch based methods as NL-means  \cite{NLmeans} or  BM3D \cite{dabov2009bm3d} and NLBayes \cite{lebrun2013nonlocal} can be easily adapted to image sequences just by extending the neighboring area to the adjacent frames.   
Kervrann and Boulanger \cite{boulanger2007space} extended the NL-means to video by growing adaptively the spatio-temporal neighborhood. Arias et al. extended the NL-Bayes \cite{lebrun2013nonlocal} to video \cite{arias2015towards, arias2017video}. Methods based in sparse decompositions are extended to image sequences  \cite{mairal2008learning, protter2009image, wen2015video, lee2015color}, as well as approaches based on low rank approximation \cite{ji2011robust, ji2010robust}.


The performance of many denoising methods is improved by  introducing motion compensation.   These compensated filters estimate explicitly the motion of the sequence and compensate the neighborhoods yielding stationary data \cite{ozkan1993adaptive}.  The BM3D extension, VBM4D \cite{maggioni2011video},  exploits the mutual similarity between 3-D spatio-temporal volumes constructed by tracking blocks along trajectories defined by the motion vectors. 
In \cite{vdenoisingTIP15} the authors proposed to combine optical flow estimation and patch based methods for denoising. 
Their algorithm tends to a fusion of the neighboring frames in the absence of occlusions and a dense temporal sampling.  In this ideal scenario, an optical flow or global registration is able to align the frames and fusion is achieved by simple averaging \cite{haro2012photographing}.  The algorithm in \cite{vdenoisingTIP15} compensates the failure of these requirements by introducing spatiotemporal patch comparison and denoising in an adapted PCA based transform.  

Recently, neural network methods appeared making use of motion compensation \cite{tassano2019dvdnet}, patch based processing  \cite{davy2019non,vaksman2021patch} or being applied recursively in groups of three consecutive frames \cite{tassano2020fastdvdnet}. 
Self-supervised methods have also been proposed by using warped neighboring frames to define the loss function \cite{dewil2021self}.

\subsection{Noise removal at the CFA and joint denoising-demosaicking}

Since demosaicking is the main cause of noise correlation, it is suitable to remove the noise before this process, or simply combine them in a single procedure. 
Paliy et al. \cite{paliy2008denoising} performs interpolation and denoising of a single CFA Bayer image. 
The method uses inter color filters selected by polynomial approximation.   
Chatterjee et al. \cite{chatterjee2011noise}  denoises the CFA by using the NL-means algorithm. 
Their method averages only patches  having the same CFA pattern. The authors do not use any variance stabilization transform.  They propose a demosaicking method based on an optimization process where the CFA is taken as the low resolution counterpart in a super resolution framework.

Zhang \etal\cite{zhang2010spatial} proposed to first denoise the CFA before applying a spatio-temporal demosaicking algorithm. A spatio temporal extension of \cite{zhang2010two} is applied for denoising by combining only patches having the same CFA pattern. Noise is reduced by thresholding in an adaptive PCA basis.   An initial demosaicking is processed by a spatial-temporal algorithm in order to reduce artifacts.  For a given pixel to be processed,  one searches for the similar pixels to it within the spatial-temporal neighborhood and then let the enhanced pixel be the weighted average of them.

Heide et al. \cite{heide2014flexisp}  proposed a single optimization process being able to deal with all image chain stages.  Denoising, demosaicking and deconvolution are written in a single energy minimization solved by primal-dual techniques \cite{chambolle2011first}. The method applies to a single image, even it can be adapted to deal with image sequences. Patil et al. \cite{patil2016poisson} denoise the CFA sensor data by a dictionary learning combined with a variance stabilization transform.  No particular demosaicking algorithm is proposed.
Hasinoff et al. \cite{hasinoff2016burst} proposed a chain for burst sequences of images, with the novelty of dealing with the Bayer data. The method aligns and merges all images of the burst sequence into a single one.  Subsampled Bayer of factor 2 is used for alignment, a tiled translation is estimated by a Gaussian pyramid process.  Finally, a robust merging method is implemented using the FFT.  

{ Neural networks have also been recently proposed for burst fusion Mildenhall et al. \cite{mildenhall2017burst} and joint denoising and demosaicking \cite{kokkinos2019iterative}. }

\subsection{HDR static}

Classical HDR methods proposed several weight functions and CRF estimation strategies to be applied with the averaging  \eqref{eq:fusion_hdr}.  
Mann and Picard \cite{mann1994beingundigital} proposed to use as weighting function the derivative of the CRF, since greater response sensitivity corresponds to greater certainty on the irradiance values. Devebec and Malik \cite{debevec2008recovering} and Khan et al. \cite{khan2006ghost} use both a hat function, based on the assumption that well exposed values, closer to the mid-range, are more reliable than over or under-exposed pixels. Based on signal theory, Mitsunaga and Nayar \cite{mitsunaga1999radiometric} propose to multiply Mann and Picard’s weight by the response output, since larger values are less influenced by a constant noise floor. Figure \ref{fig:hdr_weights} shows some example of weighting functions for HDR fusion for static sequences. 

More recent methods are proposed to perform HDR fusion by using deep learning techniques, such as classic convolutional neural networks \cite{kalantari2017deep} or other models, such as generative adversarial networks \cite{niu2021hdr}. Although these methods aim at improving the visual quality of the result, they are more devoted to remove ghosting artefacts due to motion of the camera or objects in the scene.

\subsection{HDR deghosting}

Tursun et al. \cite{tursun2015state} presented a taxonomy for classification of HDR deghosting algorithms. 
Here we provide a summary of the proposed classification, as well as more recent methods posterior to the published taxonomy. 
For a more extense review, please refer to Tursun et al. article. 

\medskip

The proposed classification divides the methods into two groups: on one hand, methods that remove artifacts due to global camera motion (named global alignment algorithms) and methods that remove artefacts that result from moving objects in the scene (named moving object algorithms). This second family of methods usually assume that the camera is static. Both families can be combined for hand-held camera deghosting.

\medskip

Global alignment algorithms aim at detecting different global motion models: translation \cite{ward2003fast, akyuz2011photographically}, rotation \cite{cerman2006exposure}, affinity \cite{im2011geometrical} or homography \cite{mann2002painting, tomaszewska2007image}.  There exist a variety of methods for estimating the desired motion model: Rad et al. \cite{rad2007multidimensional} and Yao \cite{yao2011robust} propose to register the images by translating the problem to the Fourier domain; Mann et al. \cite{mann2002painting} and Candocia \cite{candocia2003simultaneous} use the comparametric equations to find the desired global transformation; Tomaszewska and Mantiuk \cite{tomaszewska2007image} and Gevrekci and Gunturk \cite{gevrekci2007geometric} find the transformation by matching image descriptors, such as SIFT \cite{lowe2004sift} or CIFT.


\medskip

Moving object algorithms can be divided in several subclasses, depending on the main strategy used to remove ghosting artefacts: moving object removal, moving object selection or moving object registration. Moving object removal algorithms replace selected regions by an estimation of the static background. 
These methods may fail in presence of dynamic backgrounds, moving objects with occlusions or insufficient number of exposures \cite{khan2006ghost,pedone2008constrain,granados2008background,silk2012fast}.
Moving object selection detects the presence of moving regions and substitute those pixels by corresponding areas from other images \cite{grosch2006fast,jacobs2008automatic,raman2009blind}. 
Moving object registration methods remove the majority of artefacts by computing motion between a reference frame and the other images of the stack. The methods in this family can also be classified in two subclasses: optical flow methods \cite{hossain2011high, zimmer2011freehand, FerradansBPC12} or patch based methods \cite{hu2012exposure, sen2012robust}. The methods in the first group aim at finding a pixel-wise motion between images, while the ones in the latter use patch-based strategies to register images and remove artefacts. 

\medskip

{
There also exist more recent methods using deep learning techniques to remove ghosting artefacts.
The first method using a convolutional neural network was Kalantari et al. \cite{kalantari2017deep}, which perform a pixel-wise merging. 
Wu et al. \cite{wu2018deep} propose a non-flow-based approach for HDR; Yan et al. \cite{yan2019attention} handle motion between images by an attention module and later they proposed a non-local network \cite{yan2020deep}. HDR-GAN \cite{niu2021hdr} propose a novel GAN-based model with a novel generator network. This network contains a reference-based residual merging block, which is able to align large objects and camera registration. Other artefacts are removed by a deep HDR supervision scheme. Liu et al. \cite{liu2021adnet} align the dynamic frames with a deformable alignment module. They propose to use a two branch network, to process separately LDR images and irradiance images, since they provide different information about the scene.} 

\subsection{Joint HDR and noise removal}

There is very few literature of HDR imaging dealing with noise removal.
Akyuz et al. \cite{akyuz2007noise} denoise each frame before fusion, by averaging in the radiance domain a subset of frames. 
Tico et al. \cite{tico2010motion} combine an initial fusion with the { image of the sequence with the shortest exposure} in the luminance domain.
This combination is performed in the wavelet domain and coefficient attenuation is applied to the coefficients of the difference of luminance{s}. Min et al. \cite{min2011noise}  filter the set of images by spatio-temporal motion compensated anisotropic filters prior to HDR reconstruction. Lee et al. \cite{lee2014ghost} use sub-band architecture for fusion, with a weighted combination using a motion indicator function to avoid ghosting effects.  The low frequency bands are filtered with a multi-resolution bilateral filter while the high frequency bands are filtered by soft thresholding. Ahmad et al. \cite{ahmad2016noise}  identify noisy pixels and reduce their weight during image fusion. 
Goossens et al. \cite{goossens2012realistic} propose a realistic noise model for HDR imaging that takes into account an accurate noise model of image capture. 
With that, they modify the HDR weights to improve the PSNR. 
Kronander et al. \cite{kronander2013unified} propose a unified framework for HDR reconstruction from raw CFA data. The proposed method is based on an adaptive spatial and cross-sensor filtering using a polynomial approximation. During reconstruction, they perform CFA interpolation, resampling and HDR assembly in a single operation.

\subsection{Tone Mapping}



Tone-mapping algorithms are closely related with HDR imaging since they are necessary for the visualization of HDR images in common display devices. 
Tone-mapping algorithms  can be classified into global and local methods. 
Global methods apply a single tone-mapping curve on each pixel of  the image \cite{tumblin1993tone, ward1994contrast}, while local methods make use of local spatial properties to perform the task adaptatively on each pixel \cite{gu2012local}.
Global methods require less computation and are faster than local ones. 
Durand and Dorsey \cite{durand2002fast} proposed a bilateral filter which is able to preserve edges and remove most halo artefacts. 
Mantiuk et al. \cite{mantiuk2006perceptual} proposed a contrast processing framework. 
Farbman et al. \cite{farbman2008edge} use a multi-scale strategy and a least-square filter to perform tone-mapping. 
Paris et al. \cite{paris2011local} proposed a local Laplacian filter which increases contrast locally by applying a monotonic remapping function to the coefficients of a Laplacian pyramid. 

Although classical methods for tone-mapping produce good results, hyper-parameter tuning is needed to achieve the best visual quality and  reduce halo effects. To overcome these problems, deep learning-based methods were proposed: a generative adversarial network (GAN) \cite{patel2017generative}, an autoencoder network with skip connections \cite{yang2018image}, conditional generative adversarial networks (cGAN) \cite{rana2019deep} or bicyleGAN \cite{su2021explorable}.

\section{Proposed joint HDR and noise removal method}\label{sec:method}

The proposed algorithm performs a joint denoising and HDR of a reference RAW image given a series of---not necessarily static---accompanying images taken under a variety of exposure times.

\subsection{Noise model with different exposure times}

Let $\mathcal{I} = \{ I_1, I_2, \dotsc, I_N\}$ be a sequence of noisy RAW images with corresponding exposure times $\{ \tau_1, \tau_2, \dotsc, \tau_N\}$.  As in ~\cite{Buades2020CFA}, each one of these images is converted into a 4-channel image of half the width and height of the original RAW one, containing the red, blue and two green values. These are still denoted as $I_i$.

\medskip

The first step we take is to estimate the level of noise that is present in each one of the four channels of the disassembled CFA. Since all images are assumed to have been taken with the same ISO value, their noise curves $\{ \sigma_1, \sigma_2, \dotsc, \sigma_N\}$ are identical and denoted by $\sigma(x)$, depending on the color value $x$.

\medskip

We use the same approach in \cite{Buades2020CFA} which adapts the single image noise estimation algorithm in \cite{colom2014nonparametric}. This algorithm divides each image in patches and applies the DCT as proposed by Ponomarenko \cite{ponomarenkoLZKA07} for uniform noise estimation.  The low frequencies of the DCT permit to select the less oscillating patches, and the high frequencies of these selected patches yield a standard deviation estimate.

\medskip

Patches from all the initial images are classified in bins depending on its mean, which permits to estimate a standard deviation for each color, and thus an intensity dependent model.  The algorithm yields a set of observations $\{ x_i, \sigma(x_i) \}$ which has to be interpolated to the whole range in order to have a complete noise model.  Noise at the sensor is often assumed to follow a Poisson distribution, with linear variance. 

The noise estimation is applied independently for each of the four channels, leading to the estimation model $\sigma=(\sigma_r, \sigma_{g_1}, \sigma_{g_2}, \sigma_b)$.

\subsection{Normalization and variance stabilization}

Let  $I_{\text{ref}} \in \mathcal{I}$ be chosen as reference, we perform an initial normalization step by equalizing the exposure of each image to this reference one. This is accomplished by normalizing the exposure time taking care of the black offset, 
\begin{equation}  \hat{I}_i = O+ \frac{\tau_{\text{ref}}}{\tau_i} (I_i - O), \label{eq:normalization}\end{equation}
being $O$ the constant black offset image. The image $I_{\text{ref}}$ is chosen to be the middle-exposed one.  Due to this normalization, the noise model of each $\hat{I}_i$  becomes
\begin{equation}
\label{eq:noisescaled}
\bar{\sigma}^2_i(x) = \frac{\tau^2_{\text{ref}}}{\tau_i^2} \sigma^2\left(\frac{\tau_i}{\tau_{\text{ref}}} x\right) \quad \forall i \in \{1, \dotsc, N\}.
\end{equation}

\medskip

Given any image with signal dependent noise model $\sigma(x)$, a variance stabilization transform $f(x)$ permits to convert this image into a new one having a uniform noise variance,
$$f_k(x) = \int_0^x \frac{\sigma_0 \, dt}{\sigma_k(t)},$$ 
being $\sigma_0$ a fixed noise standard deviation and $k$ indicates the image channel. When assuming a linear variance model, this stabilization is known as the Anscombe transform.  This transformation is different for each channel, since each one has its own noise function $\sigma=(\sigma_r, \sigma_{g_1}, \sigma_{g_2}, \sigma_b)$.

We apply such a variance stabilization transform to each one of the normalized images $\hat{I_i}$.  We use the same esimated noise curve to modify every $\hat{I}_i$. It can be proved that, if the noise variance is approximated linearly, the uniform variance of the stabilized images, $\hat{\sigma}_i$, differs from $\sigma_0$ by exactly a scaling factor $\sqrt{\tau_{\text{ref}} / \tau_i}$ and $\hat{\sigma}_\text{ref}=\sigma_0$.

\medskip

The color of the variance stabilized images is comparable even if their noise standard deviation is different.  We will take advantage of that in the next step, where the data from brighter, less noisy images will have a greater impact on the final~result.

\subsection{Joint denoising and HDR}

From now on, we assume that every frame $\hat{I}_i \in \hat{\mathcal{I}}$ has a uniform noise standard deviation $\hat{\sigma}_i$, while any mention of $I_i \in \mathcal{I}$ refers to the original, pre-normalization image using the RGGB space.

\medskip

We will use a decorrelation transform YUVW for each 4-channel image $\hat{I}_i$ given by matrix 
$$M=\left(\begin{array}{cccc} 0.5 & 0.5 & 0.5 & 0.5\\ -0.5 & 0.5& 0.5& -0.5 \\ 0.65 & 0.2784 & -0.2784&-0.65 \\-0.2784  & 0.65& -0.65& 0.2784 \\ \end{array}\right).$$
This transform was computed by applying a Principal Component Analysis (PCA) to several 4-channel images obtained from RAW.
If we assume that noise values at the sensor data are uncorrelated for different color channels and pixel locations, since this is an orthonormal matrix, its application to $\hat{I}_i$ conserves the decorrelation and uniform standard deviation properties. The transposed matrix yields the inverse transform.

\medskip

Firstly, we compute the optical flow between the better exposed frame $\hat{I}_{\text{ref}}$ and every other $\hat{I}_i \in \hat{\mathcal{I}}$ on the brightness channel Y.
We check the consistency of the obtained values by also computing the inverse flows and making sure that they are approximately reciprocal.
Pixels that do not verify this condition are either occluded or in violation of the optical flow's colour constancy assumption, so we mark them as invalid for the rest of the algorithm. { Then, we warp each frame onto the reference image; at this point we have a sequence of static images along with their respective occlusion masks.}

\medskip

Afterwards, a 3D volumetric patch-based approach is used to search for similar patches, while still 2D image patches are used for denoising and HDR synthesis.  For every overlapping $k \times k$ patch in $\hat{I}_{\text{ref}}$, $P$,  the  patch ${\cal P}$ referring to its extension to the temporal dimension is considered, having $N$ times more pixels than the original one. Since the images have been resampled according to the estimated flow,   the data is supposed to be static.  The algorithm looks for the $K$  extended patches closest  to ${\cal P}$.  
A matrix containing the pixel values for its $K$ most resembling extended patches of size $N$ (i.e.\ a matrix of dimension $KN \times k^2$) is built.

\medskip

Let $P$ be an arbitrary patch in $\hat{I}_{\text{ref}}$, and $\mathbf{X}$ the matrix whose $j$-th row is composed of the flattened pixel values of the 2D patch $Q_j$. We wish to obtain from $\mathbf{X}$ a HDR and noise free patch.  To do so, we carry out a Weighted Principal Component Analysis. For every neighbouring patch $Q_j$ of $P$, we compute its associated weight
\begin{equation}
\label{eq:weights}
w_j = w_{\text{sim}}(P, Q_j) \cdot w_{\text{HDR}}(Q_j) \cdot w_{\text{SNR}}(Q_j),
\end{equation}
where
\begin{itemize}
  \item $w_{\text{sim}}(P, Q_j) = \exp\left( - \|P - Q_j\|^2 / (h\, \hat{\sigma}_{\text{ref}})^2 \right)$ weights $Q_j$ so that patches that are too dissimilar to $P$ lose influence on the filtering step (depending on a parameter $h$).
  \item $w_{\text{HDR}}(Q_j)$ is any HDR weighting function adapted to the RAW range of intensities,  evaluated on the same position and original RGGB image $I_i$ where $Q_j$ is located.
  \item $w_{\text{SNR}}(Q_j) = \sqrt{\tau_i / \tau_{\text{ref}}}$ compensates for the fact that $\hat{\sigma}_i \neq \hat{\sigma}_{\text{ref}}$, favouring patches $Q_j$ from a frame $i$ with a higher exposure time and thus better signal to noise ratio after normalization and variance stabilization.
\end{itemize}
The weights $\{ w_j \}$ are then used to center the data in $\mathbf{X}$.
That is, we subtract the vector
\[
\mathbf{b} = \frac{1}{\sum_{j=0}^{KN} w_j} \sum_{j=0}^{KN} w_j \mathbf{x}_j
\]
to each row $\mathbf{x}_j$ of $\mathbf{X}$ to form the centered data matrix $\bar{\mathbf{X}}$.  After that, we obtain the eigendecomposition of the weighted covariance matrix
\[
\frac{V_1}{V_1^2 - V_2} \bar{\mathbf{X}}^T\mathbf{W}\bar{\mathbf{X}} = \mathbf{V}\mathbf{S}^2\mathbf{V}^T,
\]
where $\mathbf{W} = \operatorname{diag}(w_1, \dotsc, w_{KN})$, and $V_1$ and $V_2$ are the sum of the weights and the sum of their squares, respectively.
This operation can be efficiently done by computing the SVD of $\mathbf{W}^{\frac{1}{2}}\bar{\mathbf{X}}$.
The matrix $\bar{\mathbf{X}}$ is then filtered by applying a threshold on the coefficients of the PCA based on the magnitude of their associated principal values compared to the expected noise, and the patch $P$ is subsequently reconstructed by adding back $\mathbf{b}$ to its corresponding row in the matrix.
This process is done independently for each channel and each patch on the reference frame, and an image is formed by aggregating those patches.
Lastly, we apply a color space conversion back to RGGB from YUVW and we undo the variance stabilizing transform to obtain the resulting~image.

{In particular, we note that, for noise-free images, using a window size of $k=1$ and setting $w_{\text{sim}} = w_{\text{SNR}} = 1$ with a full cancelation of coefficients of the PCA---that is, keeping only the barycenter $\mathbf{b}$---is equivalent to a classic HDR procedure with weights $w_{\text{HDR}}$. Indeed,  the vector $\mathbf{b}$ is a spatio-temporal generalization of the classical HDR averaging with additional noise removal when using all the weight factors. The weight $w_{\text{sim}}(P, Q_j) $ permits the noise removal but also avoids creating ghosting artefacts. 
While the barycenter estimation would be enough to remove noise, increase dynamic range and avoid ghosting effects, texture and details might be over-smoothed.  The use of PCA avoids such a detail smoothing and improves the deghosting capabilities of the method. }

\subsection{Imaging chain}

HDR methods that are applied to common 8-bit images require a secondary step to transform the values of the radiance map that they yield to an image that can be viewed on a LDR display.
This is also de case for the RAW HDR image that our algorithm outputs.
However, since a radiance map and a RAW HDR image are fundamentally different, an extra set of actions need to be taken to correctly visualize the~latter.

First of all, the denoised and HDR image that results from the procedure still has the structure of a 4-channel disassembled CFA.
For that reason, we must reassemble it and interpolate its values into the usual red, green and blue chromatic components that can be perceived by the human eye.
We accomplish this by demosaicking the image, for which we can use any state of the art method, such as \cite[Section II]{duran2014self}.

After demosaicking, the signal strengths of the three colors can be somewhat unbalanced due to how the camera sensors read different types of light.
Therefore, to conform them to a more realistic hue, a white balancing step is performed by multiplying each channel by a different value---chosen depending on the processed image.
Moreover, to be able to visualize the image on a standard display, we apply a linear transformation to turn the camera RGB values into coordinates in the sRGB color space.

Finally, we adjust the contrast and brightness of the HDR image and compress its range so that it fits in 8 bits per channel.
While a simple scaling and gamma correction is usually good enough to do so on non-HDR RAW images, we have found that more complex tone mapping algorithms \cite{paris2011local} produce more pleasant results.
Some of these algorithms have the disadvantage that they magnify image noise considerably; this is not a problem for us, given that our images have already had their noise removed.

\section{Experimentation}\label{sec:experimentation}

We evaluate the HDR capability of our algorithm by testing it on sequences of real images with different exposure times and some amount of movement between frames.  We study the effects of the different weights and the type of filtering in the proposed scheme.

The same parameters have been used in all experiments.  We keep a fixed window size of $k=7$, a parameter $h=2.0$, a threshold of principal values $\tau=2.8$. The HDR weighting function is 
$$w_{i}(P) = 1 - \left( 2 \cdot \frac{I_{i}(P) - O}{M - O} - 1 \right)^{12}$$
$$w_{HDR}(P) = (w_r(P) + w_{g1}(P) + w_{g2}(P) + w_b(P)) / 4$$
being $O$ the black offset value, $M$ the RAW maximum value ($M=4095$ for current examples) and $I_i(P)$ the mean value of channel $i$ in patch P. 
The resulting images are processed with Matlab's localtonemap function, which implements the tone mapping algorithm in \cite{paris2011local}.

\begin{figure}  \graphicspath{{figuresHDR/}}
\centering
\includegraphics[width=\textwidth]{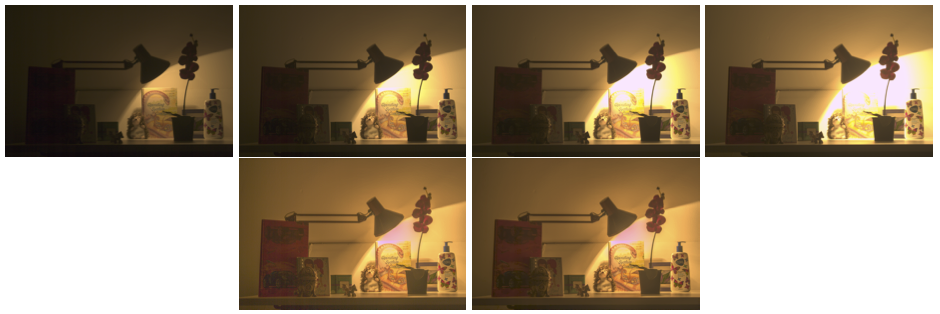}
\caption{Comparison of proposed scheme to classical HDR with the same HDR weighting function. Top: initial low dynamic images. Below, from left to right:  classical HDR and  proposed scheme. } \label{fig:example1}
\end{figure}

The RAW images in Figure~\ref{fig:deghosting} was acquired by ourselves while the images in Figures~\ref{fig:setup2denoising}-\ref{fig:example1} were taken from \cite{karaduzovic2017multi, karadjuzovic2017assessment}.

\subsection{HDR in real noisy images}

\begin{figure}
\centering
\includegraphics[width=\textwidth]{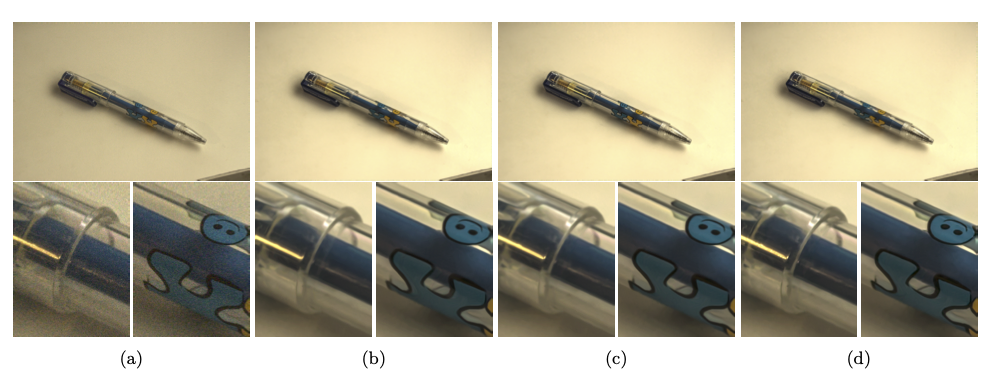}
\caption{Ghosting effects of the HDR process are corrected with the introduction of similarity weights, and details are further improved when keeping or canceling PCA coefficients based on the noise standard deviation. (a) Tone mapped reference noisy image. (b) HDR without similarity weights (only barycenter centering). (c) HDR with similarity weights (only barycenter centering). (d) HDR with similarity weights (filtering of PCA~coefficients).}
\label{fig:deghosting}
\end{figure}

First, we set the method to keep none of the coefficients of the PCA, so that it only centers the patch values to those of the weighted barycenter.
That way, we can study the effect each separate type of weight has on the HDR process.
In particular, we notice that the introduction of patch similarity weights helps against the presence of ghosting that usually appears when applying HDR algorithms on moving images.
This strong point arises from the fact that a weighting based solely on pixel intensities can mistakenly merge patches that have been aligned inaccurately; those patches give up their influence when attaching an extra weight that compares them to the reference patch.
This deghosting effect is made clear in Figure~\ref{fig:deghosting}: the translucid edges of the pen that bleed through its interior in subfigure 2b disappear completely in subfigure 2c.

\begin{figure}
\centering
\includegraphics[width=\textwidth]{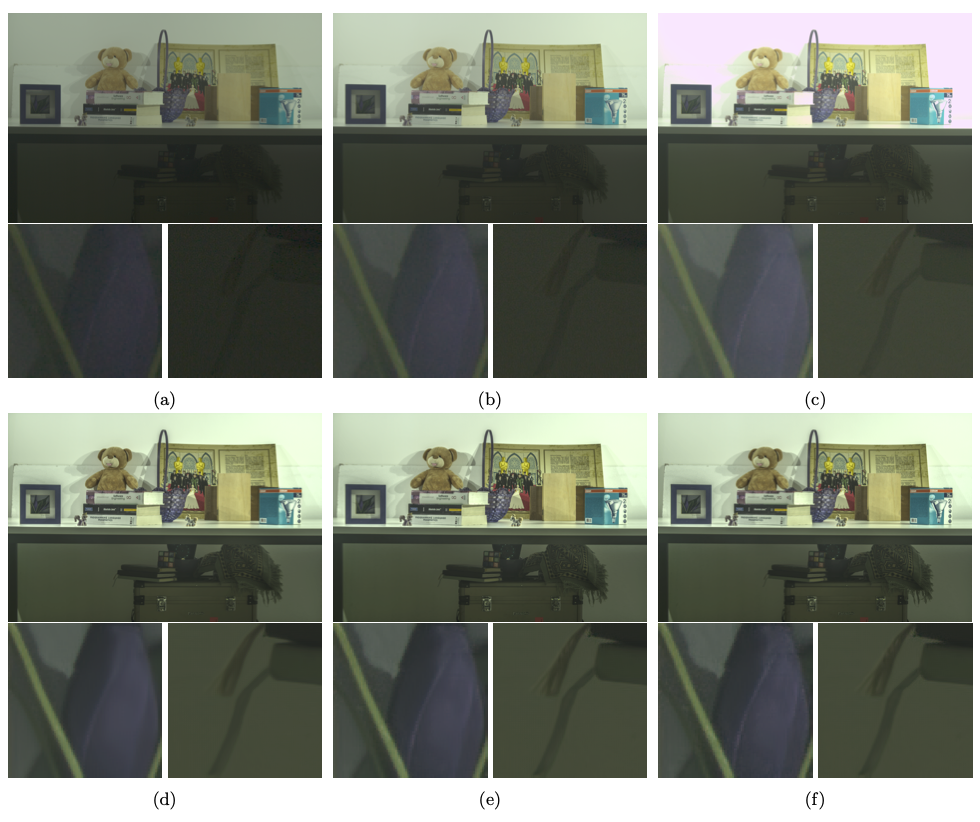}
\caption{Weighting the PCA allows us to keep fewer coefficients without compromising too much quality. (a)--(c): original images with different exposure times. (d) HDR only with barycenter centering. (e) HDR with filtering based on noise variance (at most 3 coefficients). (f) HDR with filtering based on noise variance (no coefficient restriction).}
\label{fig:setup2denoising}
\end{figure}

\medskip

We also study how the type of filtering affects both the denoising and the HDR quality.
It is clear that keeping only the barycenter of each group of patches oversmoothes the results, and keeping or canceling PCA coefficients based on the expected quantity of noise shows a higher level of detail.
We have found that, when computing a weighted SVD, it is usually enough to keep a lower number of coefficients than in the weightless case; that is, the detail information is shifted to gather on the first few components of the PCA.
In view of that, retaining PCA coefficients up to certain point adds texture without compromising the quality of the HDR.
Past that point, almost only noise is added.
For that reason, it can be beneficial to limit the maximum number of coefficients to keep, even if the filtering is done comparing the singular values to the noise variance, to make sure that no residual noise is accidentally preserved while also saving some computation time. 
Figure~\ref{fig:setup2denoising} demonstrates how our algorithm correctly denoises and expands the dynamic range of images with very dark and bright regions: comparing subfigures~3d, 3e and~3f, it is apparent that a small number of extra PCA coefficients are sufficient to add considerable detail to the resulting image with respect to performing a simple weighted average.



%

\bibliographystyle{plain}
\bibliography{references,references2,refs, references_marco}

\end{document}